# Recombination of atomic Hydrogen physisorbed on low-temperature Surfaces

Thomas R. Govers*

Dipartimento di Fisica, Università di Trento, Via Sommarive 14, I-3050 Povo,TN, Italy

* Permanent address: Aecono Consulting, 59, rue de Prony, 75017, France;
E-mail: thomas.govers@wanadoo.fr

PACS: 34.50 Dy; 95.30 Dr; 95.30 Ft

## Abstract

Molecular beam experiments that use low-temperature bolometers as (energy-) detectors are well suited to the study of physisorption and recombination of hydrogen on low-temperature surfaces. Experiments where this technique is combined with mass spectrometry to examine atoms and molecules released from the surface are summarised and reviewed with reference to astrophysical implications. Hydrogen atoms physisorbed on polycrystalline water ice are shown to be sufficiently mobile to recombine efficiently even at surface temperatures as low as 3 K. Molecules are formed with substantial internal energy, probably of the order of 35000 K, and are immediately released when formed. Coverage by molecular hydrogen plays an important role in determining overall recombination efficiency and may self-regulate recombination in interstellar clouds: on hydrogen-free grains recombination is limited by the low sticking coefficient of hydrogen atoms, while on grains covered by molecular hydrogen the binding energy is reduced so that recombination is limited by the rapid evaporation of physisorbed atoms.

## 1. Introduction

More than 50 years ago, van de Hulst [1] proposed that chemical reactions on dust grains could lead to the formation of a variety of molecules in interstellar space. Surface chemistry on low-temperature surfaces is particularly relevant to the formation of molecular hydrogen in the interstellar medium, since in most regions of space other mechanisms for recombination of hydrogen atoms do not appear to be sufficiently effective. The literature on this subject is abundant, with theoretical papers far outnumbering experimental investigations. Overviews of the basic concepts and many important early references can be found in [2] and [3], while [4] and [5] refer to more recent investigations. Yet, fascinating questions continue to be subject of discussion, such as: can physisorbed hydrogen atoms recombine efficiently in the absence of enhanced-binding sites; should the reaction be pictured as an Eley-Rideal (ER) mechanism, or as a Langmuir-Hinshelwood (LH) process; how is the available recombination energy partitioned; what are the main surface characteristics that govern the recombination efficiency; are there significant isotope effects?

The present Comment addresses some of these questions on the basis of molecular beam experiments which employed detection by bolometry and mass spectrometry to monitor recombination of hydrogen atoms on an ice-covered substrate with temperatures between 3 and 10 K [6], [7]. The use of low-temperature bolometers for molecular beam detection was proposed by Gallinaro and pioneered at the University of Geneva in the late 1960-ies [8] [9], and the first application to hydrogen atom recombination was published in 1971 [10]. The combined bolometry/mass spectrometry work was described in the Ph.D. thesis of Lorenzo Mattera [6], but only part of it was published in more readily accessible form [7]. Summarizing and commenting the conclusions that one can draw from these experiments appeared worthwhile in view of the recent renewal of efforts, both experimentally and theoretically, to understand the title reaction in more detail.

## 2. Binding energy requirements relevant to astrophysical objects

The physical conditions in different regions of space vary considerably, and it is not obvious how recombination on grains can be effective in such a wide variety of circumstances.

Suppose, for the time being, that every hydrogen atom that strikes a grain "sticks" to the surface in the sense that it thermalizes to the surface temperature. In this optimistic view, efficient recombination on a grain simply requires that the time that any one atom remains adsorbed on the surface should be comparable to the time interval between successive impacts on the grain.

Figure 1 illustrates the time scales that should be considered. The gas temperature and density data were taken from Cazaux [11], and the time $\Delta t$ between successive impacts was calculated from the density $n_H$ and average speed $v_{av}$ of H atoms at the appropriate gas temperature:

$$\Delta t = (n_H v_{av} A/4)^{-1}, \quad \text{(Eq. 1)}$$

where $A$ is the grain surface accessible to impact from the surrounding gas. In the figure, the surface considered is that of a sphere of 0.1 μm diameter. Grains with a "rough" surface will have an accessible surface larger than that of a smooth sphere, but as we are considering the "first strike" surface seen by the incoming gas atoms, the order of magnitude should be similar.

Equating these impact intervals to required residence times will yield lower limits to the binding energies compatible with efficient surface recombination, since we are assuming that every atom that remains adsorbed long enough will recombine when a second atom strikes the grain. As illustrated by fig. 1, the minimum required residence times can be quite different, depending on the region of space considered. In the diffuse interstellar clouds, with $T_{gas}$=100 K and H densities ranging from 1 to 100 cm$^{-3}$, the required residence time on a 0.1 μm grain lies between an hour and a few days. In photo destruction regions (PDR), i.e. molecular clouds irradiated by nearby stars, the values range from seconds to minutes, depending on gas temperature (10 to 100 K) and density ($n_H = 10^4$ to $10^5$ cm$^{-3}$).

These time scales are very large compared to the interval $\tau_o$ between "attempts" of an adsorbed atom to evaporate from the surface, which is typically considered to lie in the range from $10^{-13}$ and $10^{-12}$ s. We are therefore dealing with thermalized molecules, so that their residence time on the surface, $\tau$, can be described by the expression:

$$\tau = \tau_o \exp(\varepsilon/kT_s), \quad \text{(Eq. 2)}$$

where $T_s$ is the surface temperature, and $\varepsilon$ the binding energy to the surface.

It is instructive to use this expression to evaluate the binding energies that correspond to the minimum residence times illustrated by fig. 1. The answer depends on the value assigned to $\tau_o$, but the most critical parameter is the ratio $\varepsilon/kT_s$. "Sticking", in the sense of implying adsorption times described by eq. 2, can lead to hugely different residence times, depending on ($\varepsilon/kT_s$), and care should be taken as to the timescale considered when "sticking" is loosely described as "staying long". An atom that sticks to a $T_s$ = 10 K surface with a binding energy of $\varepsilon/k$ = 100 K will remain adsorbed for only $10^{-9}$ to $10^{-8}$ s.

Adopting the value $\tau_o = 10^{-13}$ s, the time constrains of fig.1 translate into the minimum binding energy requirements illustrated by fig.2. For the different regions, the figure shows the extremes corresponding to the lowest required time and grain temperature combined with the highest gas density on the one hand, or the highest time and grain temperature with the lowest gas density on the other. Energies are expressed in K (i.e. as $\varepsilon/k$; 1 eV = 11606 K), and range from about 330 K to 3400 K (28 to 290 meV) for a 0.1 μm grain. The same approach for different grain diameters yields values from 280 to 2900 K (24 to 250 meV) for 1 μm, and from 375 to 4000 K (32 to 350 meV) for 0.01 μm. The required binding energies are higher for the smaller grains because the intervals between successive impacts are longer.

The nature and characteristics of grain surfaces are not very well known. Astrophysical data and studies of meteorites suggest carbonaceous and silicate compounds as plausible materials, which, in sufficiently shielded regions, are covered with ices containing water, methanol, molecular hydrogen and heavier molecules. Uncertainties in surface composition of course imply variability in binding energies, even if the order of magnitude relevant to *physisorption* does not necessarily vary greatly with chemical composition. This is because the depth of the physisorption gas-surface potential often largely depends on the polarizability of the gas atom or molecule [12], and less on the detailed composition of the surface.

Among the best known physisorption energies are those of H and D on the (0001) surface of graphite, which Ghio et al. [13] determined from bound state resonances in atomic beam scattering. They found the binding energy for the lowest gas-surface vibration of H to be 31.6 meV (367 K) and that for D 35.4 meV (408 K), corresponding to a well depth of 43 meV (499 K). Such values are typical of physisorption of atomic hydrogen on a variety of clean amorphous or monocrystaline surfaces, as shown in the surveys by Hoinkes [12] and by Vidali et al. [14]. None of the physisorption energies listed there for H or D exceeds 50 meV (580 K).

Except for the lowest-temperature grains, binding energies of such magnitude are too small for a second atom to reach the grain with sufficient probability before the first one evaporates (see fig. 2). Efficient recombination at surface temperatures above about 10 K therefore requires that hydrogen atoms be bound with energies that are significantly larger than the physisorption energies measured for clean and smooth substrates. Interstellar grains are probably not smooth, and surface defects, impurities and porosity of adsorbed ices can provide sites with enhanced binding. As pointed out by Smoluchowski [15], too numerous strong adsorption sites are not favourable either, because they will inhibit mobility on the surface. The picture that emerges is thus that of an irregular grain surface containing sites with a range of binding energies, so that even atoms that are initially physisorbed with binding energies of ≤ 580 K will "explore" the surface and "find" a defect or impurity that provides enhanced bonding compatible with the above minimum residence time requirements.

Such a situation has been explicitly modelled by Cazaux and Thielens [4], who benchmarked their study using the thermal desorption experiments (TPD) of Vidali, Pirronello and co-workers [16]. They show that both physisorption and chemisorbed sites are required to explain why recombination remains efficient even at the higher surface temperatures, and also conclude that both thermal diffusion and tunnelling are essential to describe the mobility between sites.

## 3. Bolometer experiments on the physisorption of molecular hydrogen

According to a poetic statement taped to one of the University of Trento molecular beam machines, "a bolometer is a thermometer that can measure the heat from a polar bear's seat from a distance of thirty kilometres" [17].

In more prosaic molecular beam detection practice, it is a small slab of semiconductor- or supraconductor material maintained at sufficiently low temperature so that minute changes in temperature will significantly modify its electrical resistance [18]. The bolometer is typically mounted on the bottom of a liquid-helium cryostat, and supplied with current from a stable constant-current source. The impinging molecular beam is chopped, resulting in a temperature variation that is detected as a modulated voltage using phase-sensitive amplification. The signal is proportional to the amount of energy that the molecular beam transfers to the bolometer surface *on a timescale comparable or shorter than the chopping period*, typically 10 to 50 ms. The relation between voltage signal and energy transferred can be calibrated, so that with beams of known particle flux, it is possible to express the signal as the average energy transferred to the surface per incoming atom or molecule.

Figure 3 shows the result of an extremely simple experiment: a steady effusive beam of molecular hydrogen impacts on a 3.4 K silicon bolometer, and the signal is monitored as a function of exposure time. The surface of the bolometer is covered by polycrystalline water ice, resulting from condensation during cool-down of the liquid-helium cryostat, as outlined in the appendix. Before time zero, the beam is flagged off and the bolometer "flashed" to 20 K, to remove physisorbed hydrogen. At time zero the beam flag is opened, and the detector exposed either to the full beam or to the same beam attenuated by interposing a nominally 36% transparent mesh. In both cases, the signal increases with exposure time, with "breaks" in slope that occur at times that are 2.5 times longer for the attenuated beam. Brief insertions of the attenuation screen demonstrate that at any exposure time, the signal is linear in beam intensity. Experiments of this type are very reproducible, with closely similar results obtained in different laboratories over time spans of years.

Qualitatively, it is clear from these observations that the time evolution of fig. 3 results from the gradual increase in energy transfer efficiency as molecular hydrogen accumulates on the ice covering the bolometer surface. From the incident beam intensity and the time spans over which significant signal changes are observed, one concludes that one is dealing with hydrogen coverages of the order of at most a few monolayers.

To independently evaluate the relevant binding energies and sticking coefficients, one needs additional experimental information. With that aim, the Genoa group complemented the energy release measurements by monitoring molecules reflected or rapidly evaporated from the surface with a second bolometer [19]. At Waterloo, the molecules released by the bolometer surface were detected by a rotatable quadrupole mass spectrometer (QMS) [6] [7].

Because the latter apparatus was designed for atomic hydrogen experiments, the QMS ionizer had an open "one pass" design. Ion signals, which are proportional to densities in the ionisation region, are in such a case a measure of the ratio flux/speed, rather than of flux. If one, e.g., attempts to determine the fraction (1 - $S$) of molecules that do not stick from the ratio of reflected to incident QMS signals, one needs to account for the fact that molecules reflected from a cold surface will be slower than the incident beam, so that the QMS signal of reflected molecules will be enhanced. Any density detector has the same shortcoming and provides signals that need to be corrected for particle speed if flux information is required.

Figures 4 and 5 show simultaneous bolometer and QMS data obtained when room-temperature beams of $H_2$ or $D_2$ impinge on an initially hydrogen-free amorphous silicon bolometer held at a temperature of 3 K. The beam fluxes at the bolometer surface were $2.0 \times 10^{14}$ and $1.2 \times 10^{14}$ molecules/cm$^2$/s, for $H_2$ or $D_2$, respectively [7]. As explained in detail in ref. [7], the data were analysed by introducing three time- (i.e. coverage-) dependent parameters: the binding energy $\varepsilon$, the sticking coefficient $S$, defined as the fraction of incident molecules that thermalize to the surface temperature, and the accommodation coefficient $\alpha$, which is the fraction of the available translational and internal energy transferred by molecules that do not stick. The analysis is not unique, because the QMS "only" informs about the ratio (flux/speed), but it yields upper and lower bounds to $S$, $\alpha$ and $\varepsilon$ as a function of beam exposure time. This information is transposed into dependence on hydrogen coverage once the time (i.e. coverage-) evolution of $S$ has been quantified. The results obtained for $H_2$ and $D_2$ are shown in figures 6 and 7, respectively. On the hydrogen-free surface, the sticking coefficient is small, between 0.1 and 0.2 for $H_2$, and between 0.2 and 0.35 for $D_2$. After 10 s exposure, it has increased to about 0.4 in the first case, and to about 0.7 in the second. During this time, about $4 \times 10^{14}$ $H_2$ molecules/cm$^2$, respectively $\approx 7 \times 10^{14}$ $D_2$ molecules/cm$^2$, have accumulated on the surface. Defining a monolayer (ML) as a coverage by $10^{15}$ molecules/cm$^2$, we see that depositing between half and one ML of hydrogen on an initially hydrogen-free surface increases the sticking coefficient by a factor of two. At higher coverages, the sticking probability exceeds 0.8, while the accommodation coefficient is close to 1.

The changes in binding energy are even more pronounced. Since the sticking and accommodation coefficients rise smoothly up to exposure times of 10 to 15 seconds, the "break" which one observes in the time behaviour of the bolometer signal after about 7 s exposure (see figs 4 and 5) must be due to a rather abrupt decrease of the binding energy. At a coverage of about 0.5 ML, the binding energy is indeed found to be three to four times lower than it is on the hydrogen-free surface. This rapid decrease in $\varepsilon$ with increasing hydrogen coverage explains why, at higher surface temperatures, the bolometer signal levels of at lower asymptotes, as studied systematically by Schutte et al. [19]. When $\varepsilon$ becomes sufficiently low, even molecules that do stick evaporate too rapidly for further accumulation from the incoming beam to occur. This condition is met when:

$$\varepsilon/k = T_s \cdot \ln(N_s / \tau_0 I S), \qquad (\text{Eq. 3})$$

where $N_s$ and $I$ are the number of adsorbed atoms and the incident beam flux per unit area, $S$ the sticking coefficient, and $\tau_0$ the attempt interval of eq. 2. Using beam intensities (averaged over the 50% duty-cycle) of about $10^{14}$ cm$^{-2}$s$^{-1}$ at the bolometer surface, Schutte et al. [19] showed that when the surface temperature is raised to 10 K, the signal saturates at the first "break" in the bolometer signal. Introducing an attempt interval $\tau_0 = 10^{-13}$ s, with a coverage of $2.5 \times 10^{14}$ cm$^{-2}$ and a sticking coefficient $S = 0.3$ which apply to this situation, the corresponding $H_2$ binding energy is found to be 280 K, which fits well within the span of values of fig. 6 at the same coverage.

The dependence of the sticking- and accommodation coefficients on hydrogen coverage is not unexpected. Both parameters are a measure of the efficiency with which translational and rotational energy of the incoming molecules is transferred to lattice vibration of the substrate. Sticking occurs when the molecule is trapped in the potential well that governs the motion in the direction perpendicular to the surface [19].

When the bolometer surface is hydrogen-free, the incoming hydrogen molecules need to transfer energy to (adsorbed) molecules of relatively large mass, since the bolometer is covered by a cryodeposit formed during cool-down of the cryostat (polycrystalline water ice, see Appendix 1) or by layers of $H_2O$, Ar, $N_2$, $O_2$ or $CO_2$ purposely deposited from a secondary beam source. The unfavourable mass ratio hinders efficient energy transfer, resulting in low sticking and accommodation, with values about twice as high for $D_2$ as for $H_2$, as qualitatively expected if mass ratios are indeed of dominant importance. When physisorbed hydrogen builds up on the surface, there is an increasing probability for the incoming molecules to collide with a molecule of similar mass, and energy transfer is more efficient [19].

A quantitative discussion of the sticking of light atoms (H, D, He) on graphite (0001), based on quantum-mechanical perturbation theory, was given by Buch [20]. She concluded that inelastic gas-surface coupling is confined to highly localised spots on the surface, within a diameter of about 0.07 nm around any particular surface atom. It therefore appears reasonable, classically, to distinguish between "hitting" a preadsorbed hydrogen molecule or a heavier (adsorbed) surface molecule. Buch also showed that the sticking coefficient is very sensitive to the weak coupling between graphite layers, which implies that the "stiffness" by which pre-adsorbed molecules are coupled to the substrate should influence the sticking probability significantly.

The difference in binding energies between a surface that is hydrogen-free and one that is saturated with hydrogen is also to be expected. In the first case, the binding should be essentially that of hydrogen on water ice. As shown in figs 6 and 7, the bolometer measurements under these conditions are compatible with a range of binding energies between 360 and 650 K for $H_2$, and between 400 and 720 K for $D_2$ [7], in agreement with the 550 K calculated by Hollenbach and Salpeter [21] for $H_2/H_2O$, and the experimental value of 500 K of Yuferov and Busol [22]. Sandford and Allamandola more recently reported a value of 555 ± 35 K for binding of $H_2$ on 2:1 $H_2O$ – $CH_3OH$ ice [23]. After sufficient exposure, on the other hand, the incoming molecules "see" a surface similar to that of solid hydrogen, since the beam intensities of ref. [7] and [19] are high enough to accumulate several monolayers of hydrogen on the 3 K bolometer surface, on a time scale of tens of seconds. Asymptotically, the adsorption energy therefore approaches the hydrogen sublimation energy: 90 K for para $H_2$ and 133 K for ortho $D_2$ [24].

The rapidity of the transition from a binding energy of about 550 K to a value of about 110 K is, however, surprising. According to the data of figs 6 and 7, less than one ML is required to achieve this change. This does not seem compatible with a picture of successive monolayers covering the ice surface and reducing the binding energy to that of hydrogen on hydrogen when the substrate becomes "hidden" from the incoming gas molecules. It seems more realistic to assume that the initial ice surface presents a range of binding energies, and that the mobile hydrogen starts by occupying the more strongly-bound sites. The experiments, which provide a measure of *average* binding, start by yielding a value slanted towards the higher binding values. As the corresponding sites are progressively occupied, the average over the remaining sites becomes lower. In addition, increasing inter-adsorbate interaction must also contribute to lowering the average binding energy, since it rapidly approaches the sublimation energy of solid hydrogen. We recall that the coverage scales of figs 6 and 7 were obtained from the known incident beam flux and the geometric area of the bolometer. They are an upper limit to the coverage on molecular level, should surface rugosity or porosity be important. Since strong adsorption differences at coverages significantly lower than those indicated in figs 6 and 7 would not appear realistic, this suggests that the ice on the bolometer is relatively smooth at molecular level and/or that porous structure is not accessed at the temperatures under investigation.

Hixson et al. [25] have reported a combined spectroscopic and computational study on the adsorption of $H_2$ on amorphous ice, which was mimicked by a $(H_2O)_{450}$ cluster. Potential minima for isolated hydrogen molecules adsorbed on its surface were found to range in depth from 320 K to 1400 K, with a most probable value at 650 K. This wide range reflects the differences in coordination number when the molecules adsorb on a surface with cracks and crevices on a molecular scale. The authors also suggest that upon initial deposition, hydrogen starts by saturating the deeper binding sites. This was further shown by Buch and Devlin [26] who computed the steady-state adsorption on the cluster at different densities and temperatures. They found a binding energy as high as 737 K under conditions simulating dense interstellar clouds, while an average of 539 K applied to a simulation of laboratory experiments. The rapid lowering of the average binding energy with increasing molecular hydrogen coverage observed in the bolometer experiments is consistent with these findings.

## 4. Bolometer experiments on hydrogen atom recombination

### 4.1 Influence of molecular hydrogen coverage

Microwave or RF sources used in the hydrogen atom-beam experiments below typically operate at a temperature $T$ of 350 - 450 K. The average kinetic energy of the atoms in an effusive beam, $2kT$, is therefore between 700 and 900 K. Adding an adsorption energy of about 400 K or less (see below), the amount of energy available at the bolometer surface is thus at most 1300 K per atom, if no recombination occurs. But if all beam atoms recombine to form molecules in their lowest vibrational level, the available energy per incident atom potentially increases by 25980 K, half the hydrogen molecule's dissociation energy. If only 10% recombine, forming hydrogen in the v = 4 vibrational level, e.g., the increase will "only" be 2600 K. From the magnitude of the calibrated bolometer signal, and knowing the flux of hydrogen atoms to the surface, it is therefore possible to draw a number of conclusions about the recombination probability, but quantitatively the interpretation will depend on the amount of internal and/or translational energy assigned to the product molecules.

From the experiments of sect. 3, we expect (pre-)coverage by molecular hydrogen to have a significant effect on the probability for a hydrogen atom to stick and recombine on the bolometer surface. At high coverage, the sticking coefficient should be relatively large, but the binding energy small, so that physisorbed atoms may not remain adsorbed long enough to recombine. On a hydrogen-free surface, few atoms are expected to stick, but if they do, the binding energy should ensure long enough a residence time for recombination to occur. At intermediate coverage, a compromise may be reached between these opposing effects. We recall that these remarks pertain to a surface that binds molecular hydrogen, and by analogy also atomic hydrogen, through *physisorption*, i.e. with binding energies of less than, say, 1100 K (0.1 eV).

Figure 8, taken from Lorenzo Mattera's PhD thesis [6], supports this expectation. It shows the simultaneous bolometer and QMS signals obtained when a hydrogen atom beam (95% atoms in the beam) impinges on the bolometer at $T_s$ = 3 K, its surface having been previously saturated by $D_2$ from a secondary source. The traces represent the superposition of 50 successive runs. At the end of each, the bias current was increased to "heat" the bolometer to about 30 K – 40 K during 30 s, with both beam flags closed, in order to evaporate all physisorbed hydrogen. The bolometer temperature stabilised to its initial $T_s$ = 3 K value after about 1 min, after which the secondary $D_2$ beam was admitted, producing a signal analogous to that in fig. 5. After saturating the $D_2$ signal, the secondary beam flag was closed, and that of the primary H beam opened, resulting in signals illustrated by fig. 8. This sequence was repeated a sufficient number of times to ensure acceptable S/N for the QMS signals which were recorded, stored and averaged at the same time as the bolometer signal. The bolometer traces were highly reproducible, typically within 2 to 3% of the average shown.

At $T_s$ = 3 K, the pre-deposited $D_2$ does not spontaneously evaporate on the time scale of the successive runs, since the residence time even with $\varepsilon$ as low as 110 K (the lower asymptotic value of $\varepsilon$ in fig. 7) exceeds 10 minutes. Yet, as the surface is exposed to the hydrogen atom beam, all predeposited $D_2$ is removed within less than three minutes. Recombination locally removes pre-adsorbed hydrogen molecules from the surface.

The evolution of $D_2$ coverage was quantified as follows. After exposing the initially $D_2$-saturated surface during a chosen time to the H-atom beam, the latter was flagged off, and the secondary $D_2$ beam switched on. The $D_2$ bolometer signal was recorded until it reached its plateau value, in the same manner as shown in fig. 5. The initial value and the time evolution of this $D_2$ probe signal allowed the determination of the molecular coverage remaining when the H-beam was replaced by the secondary beam, since the correlation between $D_2$ signal and coverage is known (figs 5 and 7). Pertinent values are indicated by vertical arrows in fig. 8.

The magnitude of the bolometer peak in fig. 8 tells us that recombination is occurring, since 1800 K are transferred per incoming atom, as compared to the maximum of 1100 K available when no recombination takes place (the temperature of the microwave source was about 350 K). But the increment is small compared to the maximum 26000 K that could be released by recombination. Either recombination is inefficient, or most energy is released as internal and/or translational energy of the product molecules, or we see the combination of both.

At time zero, the atom beam of fig. 8 strikes a $D_2$-saturated surface. We know from the discussion of sect. 3 that adsorption on this surface resembles that on solid $D_2$, so that the initial adsorption energy should be well approximated by that of H on solid $D_2$. The value for adsorption on hydrogen quantum crystals was calculated by Pierre et al. [27] on the basis of two different gas-gas potentials. The lowest value found for the ground state H-$D_2$ vibration was 42 K. An experimental value of 36 K for H binding on an amorphous $H_2$ deposit was obtained by Crampton et al. [28], from which one deduces a value of 44 K for H on $D_2$ by estimating an 8 K lowering of zero point energy from the data of Pierre et al. [27].

Using the average H/$D_2$ binding energy of 43 K, the residence time for H sticking on solid $D_2$ calculated from eq. 2 is $2.3 \times 10^{-7}$ s. This is much shorter than the beam chopping period, and atoms that stick and re-evaporate so rapidly will give the same lock-in signal as non-sticking atoms with an accommodation coefficient $\alpha$ = 1. If all atoms do stick, but do not recombine, the bolometer should therefore detect an energy of $2kT$ = 700 K per atom, since the source temperature $T$ was 350 K. The actual result is 800 K, indicating that recombination is occurring, but with low probability and/or releasing little of the recombination energy to the surface.

The initial $D_2$ coverage in fig.8 is about $3 \times 10^{15} cm^{-2}$. When the bolometer signal starts rising sharply (arrow 1), it has been reduced to about $7 \times 10^{14} cm^{-2}$. At the maximum (arrow 2) it is about $2 \times 10^{14} cm^{-2}$, decreasing further to some 1.5 and $0.5 \times 10^{14} cm^{-2}$ at the times indicating by arrows 3 and 4. Throughout the coverage range, H atoms do recombine, and recombination expels $D_2$ molecules from the surface, as witnessed by the in-phase detection of mass 4 by the QMS (pre-coverage with $D_2$ rather than $H_2$ allows molecules expelled by recombination to be distinguished from recombination products). The modification in surface coverage has a large effect on the fate of incoming H atoms, as expected from the experiments with molecular hydrogen (sect. 3) and indicated by the variation in the QMS signal at mass 1. As will be quantified in the following paragraphs, the exponential increase of residence time with growing binding energy, i.e. with decreasing $D_2$ coverage, is offset by the decreasing sticking probability, and an optimum in overall recombination probability is reached at intermediate coverage, which fig. 8 suggests to lie around $2 \times 10^{14} cm^{-2}$. This is also the coverage range where one observes rapid changes in molecular binding energies (sect.3).

## 4.2 Sticking and recombination probabilities – energy partitioning

Marenco et al. [10] examined the dependence of the bolometer signal on incident beam flux both at the H-signal maximum and under hydrogen-free conditions, using a screen attenuator to vary the beam intensity. In both cases, the bolometer signal was found to be linear over the investigated flux range: from $5 \times 10^{12}$ to $10^{14}$ $cm^{-2}s^{-1}$. This means that at the same molecular coverage, the probability for hydrogen atoms to recombine is independent of the atom flux, i.e. independent of surface coverage by atomic hydrogen.

This observation implies that the Eley-Rideal mechanism makes a minor contribution to the recombination observed in the experiments under consideration. This mechanism requires a head-on collision between adsorbed- an incoming hydrogen atoms, and has recently been the subject of detailed theoretical studies [29][30][31]. Under the present conditions, the fraction of adsorption sites occupied by hydrogen atoms is of the order of $10^{-6}$ [10][6]. At such low surface coverage, the probability for a beam atom to impinge directly on an adsorbed atom is proportional to the number of surface sites occupied by H, which in turn varies linearly with beam flux. Eley-Rideal recombination would therefore result in a contribution to the bolometer signal which is quadratic in beam flux, contrary to observations.

While this does not exclude that the Eley-Rideal mechanism does participate to a minor extent, Langmuir-Hinshelwood or hot-atom recombination is apparently more important in the experiments considered. To account for a recombination signal linear in beam intensity, the low occupancy by atomic hydrogen also imposes constrains on these alternate possibilities. Mobility by tunnelling or thermal migration in the first case, or successive "hops" in the surface trajectory in the second, should ensure that before leaving the surface the incoming atom explores a number of surface sites large enough to encounter a previously adsorbed atom with a probability that depends very little on hydrogen-atom coverage.

These observations appear to be at variance with the interpretation by Katz et al. [32] of their programmed thermal desorption (PTD) measurements on HD formation on natural olivine (a silicate containing a mixture of $Mg_2SiO_4$ and $Fe_2SiO_4$) or amorphous carbon, materials considered to be representative of interstellar dust grains. In these experiments, surface temperatures were varied between 5 and 7 K, and the H and D beams that simultaneously impinged on the substrate had energies of about 200 K. It was concluded that little of the HD detected by TPD was formed during the exposure of the surface to the two atom beams, but that these molecules were mainly formed during the subsequent temperature ramp up. The low contribution of prompt reaction was considered evidence for low surface mobility of the adsorbed atoms, at least in the conditions investigated. The analysis of the data also assumed that molecules formed upon recombination do not spontaneously desorb from the surface.

In contrast, the bolometer experiments show that even at surface temperatures as low as 3 K, hydrogen atoms that stick to the surface are sufficiently mobile to recombine with a probability that can approach 1, as discussed further below. Also, recombination of physisorbed atoms not only ejects the product molecules, but in addition expels surrounding pre-adsorbed hydrogen molecules. The difference with the TPD results is to be sought in the time scales involved. Because the bolometer experiments employ phase-sensitive detection, they will directly measure only events that take place on a time scale shorter or comparable to the modulation period, typically 10 to 50 ms. This time is comparable to the residence time (in the absence of recombination) of a hydrogen atom physisorbed with a binding energy of 80 K on a 3 K surface, or with a binding energy of 240 K on a 10 K surface. As recombination of physisorbed atoms on a surface with relatively low porosity promptly ejects the product molecules formed, the timescale of minutes relevant to TPD experiments prevents these from giving direct information on the rapid reaction of the weakly bound atoms that the bolometer experiments observe. Hornekaer et al. have recently emphasised the importance of surface morphology [33], and argued that both on dense and porous amorphous solid water (ASW), hydrogen atoms are sufficiently mobile to recombine with high efficiency at 8 K, the lowest temperature accessible in their experiments. When the surface is porous, the product molecules are trapped, so that they can subsequently be observed by TPD. When the surface is non-porous, however, no product molecules are observed even when they are formed with high probability, because they are released as soon as they are formed, as concluded from the bolometer experiments.

Recent computations have illustrated the trajectories of hydrogen atoms adsorbing on amorphous clusters of water ice [34][35][36][5]. In sticking trajectories, the impinging atoms typically thermalize within a few "hops" (a few hundred femtoseconds), while backscattered atoms explore very few surface sites before leaving the surface, so that they have little chance of contributing to recombination. This means that recombination will proceed mainly by LH recombination rather than hot-atom reaction, and that it is conceptually appropriate to write the overall recombination probability of an impinging atom as:

$$P_{rec} = S.\beta, \qquad \text{Eq. 4.1}$$

where $S$ is the sticking probability, and $\beta$ the probability for a thermalized atom to recombine with another atom before evaporating. We know that, at a given surface temperature and molecular coverage, the bolometer signal, which is proportional to $I \times (S.\beta)$, is linear in beam flux, $I$. This means that $(S.\beta)$ does not depend on $I$. Since the sticking coefficient $S$ does not depend on atomic hydrogen coverage, because the latter remains very low, $\leq 10^{-6}$ ML [19][6], a linear signal implies that $\beta$ also should show little dependence on atomic coverage. This can be understood if physisorbed hydrogen atoms are sufficiently mobile to react with near-unit probability as soon as their residence time exceeds a certain critical value. The experimental value of β is then a measure of the fraction of adsorbed atoms that do meet this minimum residence time.

In order to quantify this picture, Mattera [6] analysed the simultaneous bolometer- and QMS signals of fig. 8 by extrapolating values for H binding energies as a function of $D_2$ coverage from the molecular hydrogen measurements of sect. 3. They ranged from 40 – 50 K on the $D_2$ saturated surface to 270 – 390 K on the hydrogen-free surface. The analysis was carried out by treating the internal energy of the product $H_2$, $E_{int}$, as a parameter. Two examples are shown in figs 9 and 10, where $E_{int}$ was assumed to be 15000 K and 35000 K, respectively. Comparison with fig. 8 shows that the peak in the bolometer signal observed after 90 s exposure, i.e. at a $D_2$ coverage of about $2 \times 10^{14} cm^{-2}$, corresponds to a maximum in the value of the product ($S.β$). This maximum results from the opposing trends of $S$ and $β$ with decreasing $D_2$ coverage: $S$ goes down because the H atoms have a low sticking probability on the $D_2$-free surface, while $β$ goes up as the residence time of the atoms that do stick increases. Because the relatively low bolometer signal, even at its maximum, can result both from a low value of ($S.β$) and/or from a high value of $E_{int}$, the ($S.β$) results are higher when one assumes a higher value for the parameter $E_{int}$.

The values of $E_{int}$ adopted in figs 9 and 10 should be close to the upper and lower bounds to this parameter. For $E_{int}$= 15000 K, $S$ at $t$=0 ($D_2$ saturated surface) is only 0.53, a rather low number when considering the asymptotic molecular hydrogen data discussed in sect. 3; decreasing $E_{int}$ even further would yield an unrealistically small $S_{t=0}$. With $E_{int}$= 35000 K, on the other hand, one finds $S_{t=\infty}$ = 0.16 for the sticking on a $D_2$-free surface, and a further increase would yield too high a result for the sticking on "bare" ice, in regard to the data of sect.3. These conclusions are consistent with those of Schutte et al. [19], who detected the reflected particles with a second bolometer, rather than with a QMS. They concluded that the value of ($S.β$) at its maximum should be about 0.2 for $S$ = 0.3. These are numbers similar to those of fig. 10, suggesting that the average internal energy of the product $H_2$ is indeed close to 35000 K. This means that in interstellar clouds a substantial part of the recombination energy will eventually be lost by emission from vibrationally excited $H_2$ levels, rather than contributing to collisional heating of the gas. Hornekaer et al. summarise the importance of such energy partitioning in determining the temporal and chemical evolution of molecular clouds [33].

Several theoretical calculations have become available that shed further light on these findings. The range of binding energies considered by Mattera [6] for H on the $D_2$-free surface (i.e. on polycrystalline water ice), 270 to 390 K, is compatible with the work of Buch and Zhang [34]. These authors computed minimum energies and sticking probabilities for H on an amorphous ice cluster comprising 115 water molecules. Minimum energies cover a wide range of values, most of which fall between 250 and 800 K, with a most probable value of 550 K. The sticking probability varies considerably with incident energy and is only 0.02 at 600 K, rising to 0.11 at 200 K and 0.45 at 50 K. Al-Halabi et al. [5] reported trajectory calculations for H on the (0001) face of crystalline ice, and found sticking trajectories to have binding energies in the range 400±50 K. Also in this work, the sticking probability was found to be strongly dependent on incident energy, with a 600 K value well below 0.1 at both surface temperatures investigated, 10 and 70 K. These low sticking probabilities for "fast" H atoms on hydrogen-free ice support the results of figs 9 and 10 which attribute the low recombination probability on the $D_2$-free surface to a low $S$. On the other hand, the experiments indicate that the few atoms that do stick on hydrogen-free ice recombine with high probability, which is also the conclusion of the theoretical investigations.

The internal energy of the hydrogen molecules produced by surface recombination at low temperature was briefly discussed in the seminal paper of Hollenbach and Salpeter [21]. They argued that most of the molecules resulting from recombination of mobile hydrogen atoms should be promptly released with translational and rotational energies of the order of 2000 K, leaving around 48000 K available for vibrational excitation. Several recent theoretical publications address this question in further detail. Takahashi et al. [35] used classical molecular dynamics (MD) to simulate the reaction on amorphous ice at 10 and 70 K, and found adsorbed atoms to recombine with nearly unit probability, with instantaneous ejection of the product molecules. Between 70 and 79% of the recombination energy was found to be stored in product vibration, with 10 to 15% in rotation, and 7 to 12% in translation. Only between 3 and 5% was transferred to the ice substrate. The simulation includes both LH and ER trajectories. Ko et al. [36] also used classical MD to investigate recombination of 100 K atoms on 10 K graphite, focussing on the ER mechanism. In this study too, high vibrational excitation is observed. Parneix and Bréchignac [29] have studied ER recombination on graphite at energies up to 1500 K using classical and quasi-classical trajectory calculations. Product molecules were found to recoil with substantial translational energy, close to 20000K, while about 13000 K of energy was stored in vibration. Rutigliano et al. [30] investigated ER recombination on 10 K graphite for H atoms with incident energies between 116 and 1161 K (0.01 to 0.1 eV). About 12 % of the available energy was found transferred to the substrate phonons, while the largest fraction was stored in vibrational motion. Morisset et al. used wave packet dynamics on a modelled graphite surface to study ER recombination at collision energies between 5 K and 5300 K [31], as well as LH recombination of physisorbed atoms at collision energies between 23 K and 580 K [37]. In both regimes, the reaction probability was large, except for the ER reaction at very low energies, and vibrational excitation was found to be substantial. The LH paper of Morisset et al.[37] contains a very instructive "translation" of the intricate quantum results into classical language.

The high degree of internal excitation deduced from the bolometer experiments is thus supported by several theoretical investigations. Astrophysical observations and laboratory experiments are more scarce. In spatial observation, the assignment of vibrational $H_2$ emission to excitation during the recombination processes is complicated by the extensive radiative cascades that result from UV absorption followed by fluorescence into the manifold of ro-vibrational levels of the electronic ground-state [38][39]. Emission lines that are suitable for formation diagnostics have been identified by Le Bourlot et al. [40] and include 1.733 μm emission from $v = 7$. Burton et al. [41] who observed this line in the Omega Nebula, the region of massive star formation closest to earth, have indeed concluded that it results from excitation during $H_2$ formation. Gough et al. [42] have studied the recombination of H atoms on carbon surfaces maintained at temperatures of 90 or 300 K, and observed vibrational excitation of $H_2$ up to $v = 7$ by measuring $H^-$ resulting from dissociative electron attachment: $H_2(v) + e^- \rightarrow H^- + H$. This elegant technique has the major advantage of increasing in sensitivity with increasing $v$ [42]. The dependence on surface temperature was different for the lower vibrational levels and for the higher ones, suggesting that the latter have a larger contribution from LH recombination of weakly-bound atoms. Perry and Price [43] have observed $H_2$ in $v = 1$ by rotationally resolved REMPI of hydrogen resulting from recombination on highly oriented pyrolytic graphite at 30 and 50 K. They conclude that a substantial fraction of the recombination energy is transferred to the surface. This is not necessarily in contradiction neither with the other investigations referred to above, nor with the bolometer experiments, since no REMPI information is as yet available on the vibrational levels above $v = 1$. The vibrational level closest to the value $E_{int} = 15000$ K in fig. 9 is $v = 3$, while that closest to 35000 K (fig. 10) is *v* = 7. If high vibrational levels are strongly populated compared to *v* = 1, the average energy transfer measured in the bolometer experiments may not be sensitive to that imparted by molecules in *v* = 1.

## 5. Conclusions and suggestions

Even at surface temperatures as low as 3 K, physisorbed hydrogen atoms do recombine on water ice characterised by average binding energies of about 400 K and lower. The spread in binding energies that results from (probably moderate) surface rugosity on molecular scale provides an adequate compromise between long enough residence times and sufficiently high surface mobility, so that Langmuir-Hinshelwood recombination can proceed with high probability, once atoms stick to the surface. The resulting molecules are promptly ejected, and possess a high amount of internal energy, probably close to 35000 K, on the average. Each recombination event ejects one or several hydrogen molecules that were pre-adsorbed on the surface.

While these conclusions are based on experiments on a surface covered with water ice, there is no reason why a similar behaviour should not apply to other surfaces characterised by the same range of binding energies. Indeed, bolometer experiments where a few monolayers of $H_2O$, Ar, $N_2$, $O_2$ or $CO_2$ were purposely deposited on the initial ice surface yielded results quite similar to those discussed here, both for molecular and for atomic hydrogen. In all cases, (sub-)monolayer coverage by molecular hydrogen had a very strong influence on adsorption and recombination properties.

Because of its opposing effects on sticking and residence time, molecular hydrogen can act as a moderator with respect to the surface recombination of hydrogen atoms in interstellar clouds. On grains where no molecular hydrogen is adsorbed, the overall recombination probability will be limited by the low sticking coefficient *S*, so that (S.β) should not exceed 0.2 for gas temperatures beyond about 100 K, even if β = 1. In dense clouds, on the other hand, molecular hydrogen may accumulate on the grains until it is balanced by evaporation. This typically occurs at a coverage of about 2 x $10^{14}$ $cm^{-2}$ [7], so that the overall recombination probability for atoms should be close to its optimum (S.β) ≈ 0.2.

We noted in sect. 2 that in many regions of space, residence time requirements impose relatively high physisorption energies for recombination to occur efficiently, even assuming a hydrogen atom sticking probability of 1. As illustrated by fig.2, binding energies below ≈ 1000 K are sufficient from this point of view for surface temperatures below about 20 K, and the spread in physisorption well depths for irregular surfaces should be able to match this requirement ("enhanced physisorption").

Higher physisorption energies need to be invoked to ensure long enough residence times for H atoms at higher grain temperatures, and this will also affect possible coverage by molecular hydrogen. Physisorption energies for molecular hydrogen should be higher than those of atomic hydrogen, with well depths typically larger by a factor 1.2, the ratio of the electric polarizabilities [12]. Buch and Zhang report a distribution of binding minima for H on a water cluster which peaks between 500 and 650 K [34], while Hixson et al. show a distribution for $H_2$ on a water cluster which peaks at 650 K, but which extends further to deeper potential minima [25] Considering that the decrease in zero-point energy should amplify the stronger binding for $H_2$, it seems realistic, if not conservative, to consider $H_2$ binding to be stronger by a factor 1.2 compared to H physisorption on the same surface.

This makes it possible to evaluate the binding energies for $H_2$ that correspond to the minimum H-binding requirements obtained as in sect.2. Incorporating realistic numbers for the sticking coefficients, on can then evaluate the minimum coverage by molecular hydrogen relevant to the various astrophysical regions considered. Figure 11 illustrates this for small grains in PDR clouds with "high" ($4 \times 10^4$ cm$^{-3}$) and "average" ($1 \times 10^4$ cm$^{-3}$) molecular hydrogen density, calculated with sticking coefficient $S(H) = 0.6$ and $S(H_2) = 0.7$. The H-atom density was taken to be $5 \times 10^4$ cm$^{-3}$, the gas temperature 300 K, and the grain temperature 10 K. Habart et al. have recently suggested that small-size grains (< 0.01 μm) may be the most efficient substrate for surface recombination in PDR's [44]. Figure 11 shows that if hydrogen recombination does take place on such small grains, it probably occurs on surfaces that are partially covered by molecular hydrogen. In such a situation, the probability for recombination of physisorbed H atoms will limited by the decrease in residence time resulting from the lowering in binding energy discussed in sect. 4, while "indirect chemisorption" as discussed by Habart et al. [44] may be limited by competition between H and $H_2$ for the stronger binding sites.

Deuterium molecules have higher sticking probabilities than $H_2$, especially when molecular hydrogen coverage is low, as can be seen from figs 6 and 7. Buch [20] showed that D atoms have a substantially higher probability to stick on (0001) graphite than H atoms, at least at impact energies below about 500 K. Binding energies for deuterated species are stronger, because of the lowering of the zero-point energy. Low temperature surface adsorption and recombination will therefore be more efficient for D than for H, resulting in a production ratio for deuterated molecules that is enhanced compared to the gas phase abundance. Preferential adsorption of deuterium will also enhance the formation of deuterated species which are produced by the addition, on grain surfaces, of H and D to CO, N, O and S [45]. If this difference in sticking probability is indeed as large as the factor of 2 found by Buch for energies ≤ 100 K, deuterium fractionation on grain surfaces will be significantly larger than models that assume equal sticking probabilities [46].

My last comments are a plea for a revival of - relatively simple - bolometer measurements. Clearly, residence times of hydrogen atoms are a key factor in determining the probability for recombination into $H_2$, and also for their reactions with heavier species condensed on grain surfaces. Experiments with atomic hydrogen remain quite a bit more difficult to carry out and to interpret than those with molecular hydrogen. Theoretical modelling, on the other hand, has progressed considerably. I would therefore suggest to use accurate bolometer measurements on sticking and binding of $H_2$, $D_2$ and HD to "calibrate" model calculations, and use these to predict the behaviour of atomic hydrogen under the wide variety of conditions relevant to astrophysics.

There are obvious improvements to be made when conducting bolometer experiments on molecular hydrogen. The first is to use cold beams of "normal" hydrogen and deuterium on the one hand, and cold beams of para hydrogen and ortho deuterium on the other. This allows one to calibrate the bolometer response without having to make assumptions about the accommodation of rotational energy, and once this is achieved, to study rotational energy accommodation and possibly ortho-para transitions on the surface. This could provide a valuable benchmark for computational modelling, since simulations are capable of studying the influence of molecular rotation [25][26][47]. The second improvement concerns surface preparation. It is at present uncertain to what extent the chemical composition of the interstellar grains is of importance when compared to the role played by pre-adsorbed hydrogen or by the morphology of the surface. The bolometer experiments conduct so far did not show very pronounced differences when the initial ice surface was covered by a few monolayers of $H_2O$, Ar, $N_2$, $O_2$, $CO_2$, but the experiments where not very systematic. The influence of the morphology of ice layers, which Hornekaer et al. showed to have significant effects on the retention of molecular hydrogen [33] was not investigated either. Time-resolved measurements of hydrogen molecules adsorbing and evaporating from low-temperature astrophysically relevant materials or from ice layers with different characteristics can give valuable information on the corresponding adsorption times and binding energies. Bolometers are well suited to this type of measurement, because they can be positioned close to the surface under study without having a significant effect on the sample's surface temperature. This in turn makes it possible to tune and stabilise the surface temperature quite precisely in order to achieve residence times that can be measured with sufficient accuracy.

Acknowledgments

I thank Victor Sidis and Frits de Heer for inciting me to write this paper, and Davide Bassi and his co-workers for their stimulating hospitality. Financial support by the Hydrogen Project of the University of Trento Physics Department is gratefully acknowledged.

Appendix: Characteristics of the bolometer surface

In the experiments discussed, the surface of the bolometer, which serves both as energy detector and as recombination substrate, was not well characterised by modern surface science standards. Yet, the results obtained were highly reproducible over time spans of years, from one laboratory to the other, with no significant difference between a germanium or a silicon bolometer. This reproducibility results from the similarity in procedures used to cool the bolometer to liquid-helium (LHe) temperatures. The bolometer is shielded by two concentric shrouds, cooled by LHe boiling under reduced pressure and which are surrounded by a third shroud cooled by liquid nitrogen ($LN_2$). Two series of apertures admit the beam and allow QMS detection, respectively. The vacuum chamber that houses the cryostat was evacuated by a liquid nitrogen-baffled silicon-oil diffusion pump reaching a pressure of about $10^{-7}$ Torr ($1.33 \times 10^{-5}$ Pa) with the cryostat at room temperature. Experiments start by cooling the $LN_2$ shield that surrounds the cryostat to 77 K, rapidly reducing pressure to about $2 \times 10^{-8}$ Torr. After this, the inner shields and the bolometer support are cooled, first with $LN_2$, and then with LHe. These preparation steps typically take one hour, allowing the pressure surrounding the cryostat to fall to about $10^{-8}$ Torr, while the pressure inside the cryostat should be a few orders of magnitude lower. During this procedure, the bolometer first cools down to 77 K surrounded by background gas containing mostly hydrogen from outgassing of stainless steel, and a rapidly decreasing contribution from water vapour. When LHe temperatures are reached, gas condensation on the bolometer is limited to that provided by the hydrogen beam or by gas purposely introduced from the secondary beam source, and by background gas that diffuses through the two series of apertures in the cryostat shields. Hydrogen can be removed by raising the bolometer temperature to 20 or 30 K, but this temperature is not sufficient to remove condensed water, or to induce phase transitions in the adsorbed water film.

The morphology of water films formed at low temperatures depends strongly on deposition conditions, as described by Kimmel et al. [48]. Considering the cooldown procedure described in the preceding paragraph, water is expected to condense as polycrystalline ice, since this is the structure that persists down to a temperature of about 90 K [48]. Even before the bolometer has reached this temperature, the partial pressure of water within the cryostat has become negligibly low, and further water condensation occurs from the "beams" of background gas that penetrate though the beam- and analysis apertures. We estimate the upper limit to the corresponding condensation rate to be $2 \times 10^{-6}$ ML s$^{-1}$, so that less than 0.1 ML of additional water is deposited during 12 hours of experiments. The bolometer surface viewed by the molecular beams can thus be considered to be that of polycrystalline water ice. We note that no significant modifications in the bolometer results were observed following deposition of several additional ML of water from the secondary beam source at 45° or at 90° incidence.

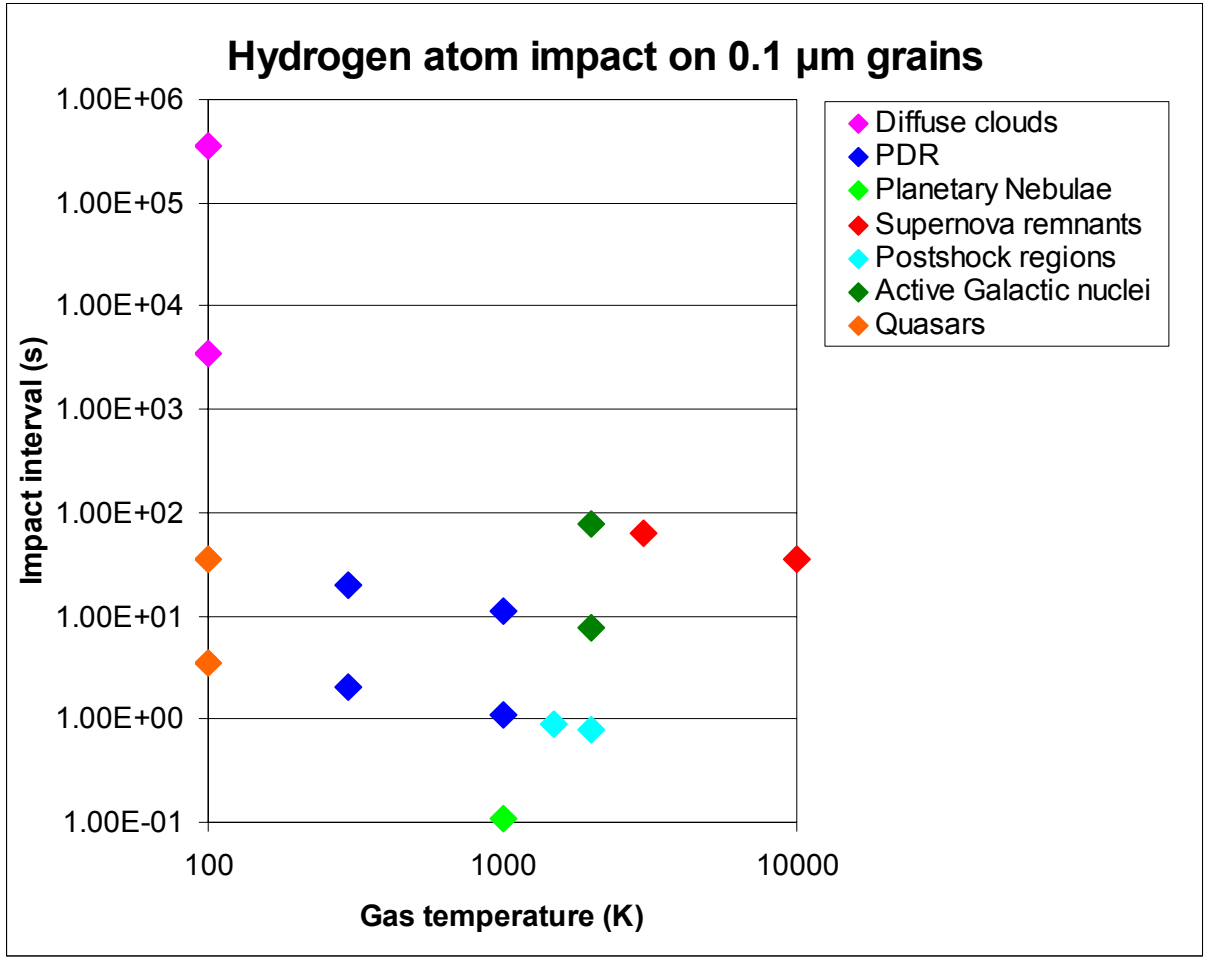

Figure 1:

Intervals between successive impacts of hydrogen atoms on a 0.1 μm grain for different astrophysical regions. Different values for the same region correspond to the range in gas density and temperature that needs to be considered. The relevant densities and temperatures were obtained from ref. [11].

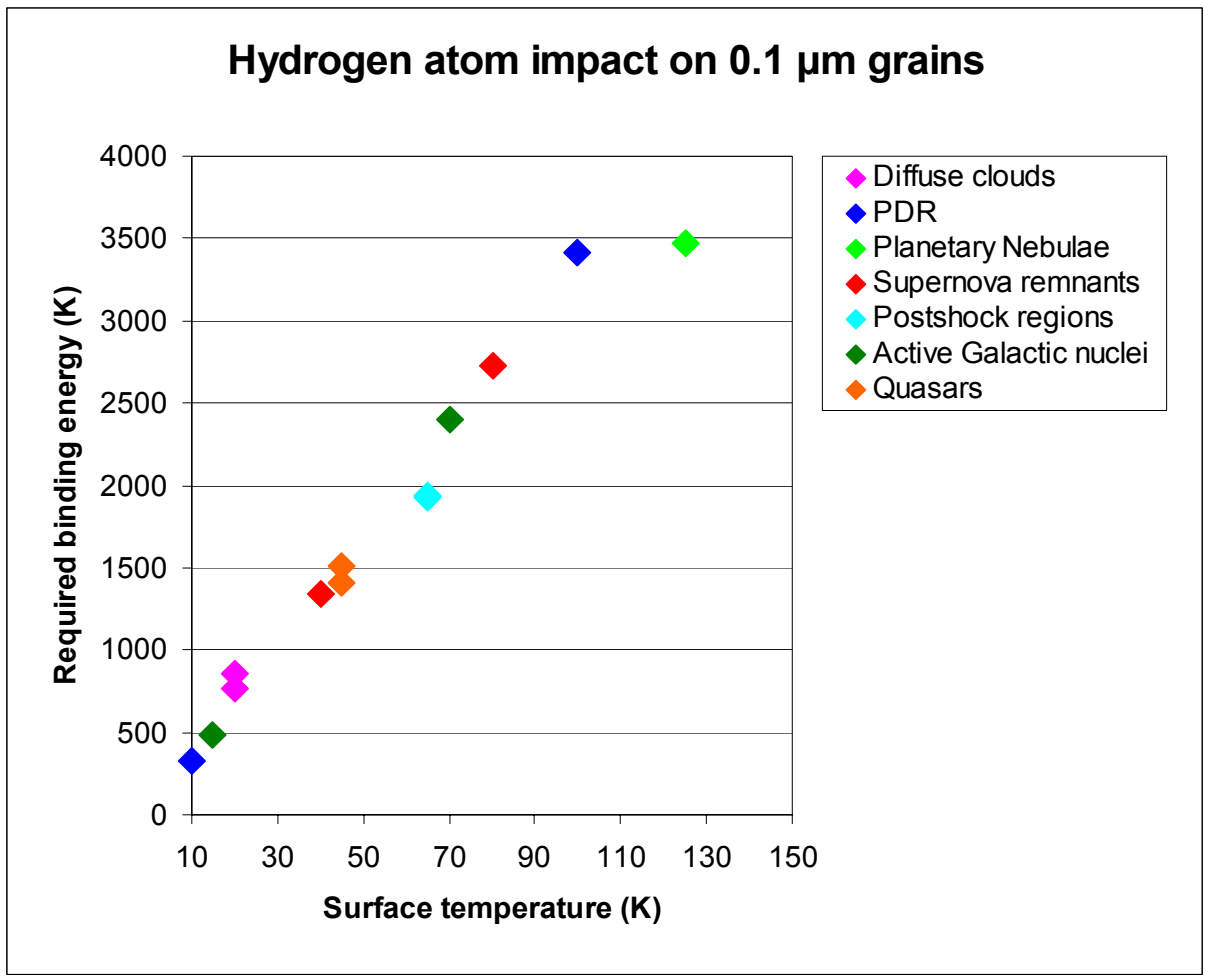

Figure 2:

Minimum binding energies required for effective surface recombination of hydrogen atoms on a 0.1 μm grain for different astrophysical regions. They were obtained by equating the impact intervals of fig. 1 to the evaporation time of a hydrogen atom thermalized to the grain temperature. Where appropriate, the figure shows the highest and lowest values corresponding to the conditions of the region considered. These were obtained from ref. [11].

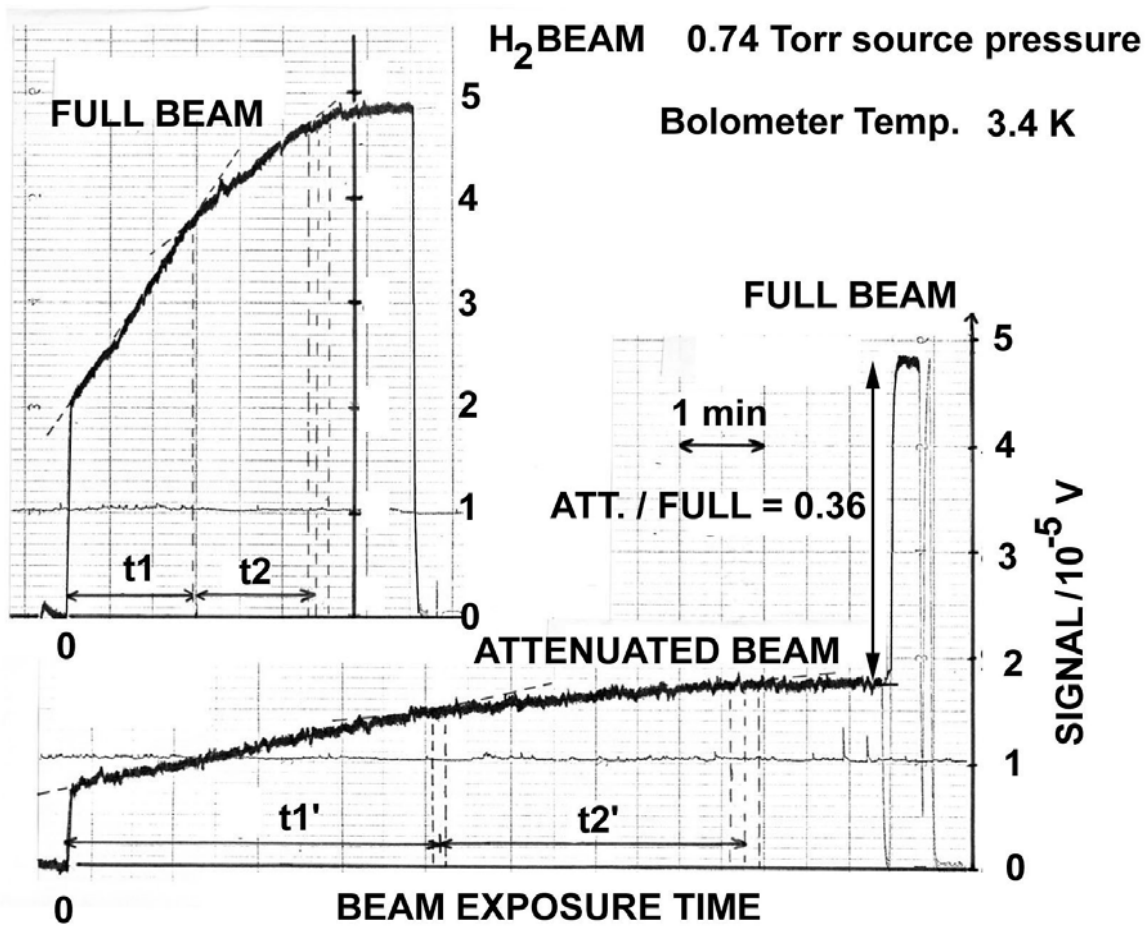

Figure 3:

Raw data showing the time evolution of the bolometer signal upon exposure to a chopped $H_2$ beam with a flux (averaged over the chopping period) of $5 \times 10^{12}$ cm$^{-2}$s$^{-1}$ ("full beam") or $2 \times 10^{12}$ cm$^{-2}$s$^{-1}$ ("attenuated beam"). Before admitting the beam, the surface has been "flashed" to remove adsorbed hydrogen. A first discontinuity ("break") is observed after times $t_1$ and $t_1'$, while a second change in slope is reached after times $t_2$ and $t_2'$. Intensities and characteristic times are observed to scale with the 36% transparency of the screen used to attenuate the beam. The bolometer temperature is 3.4 K.

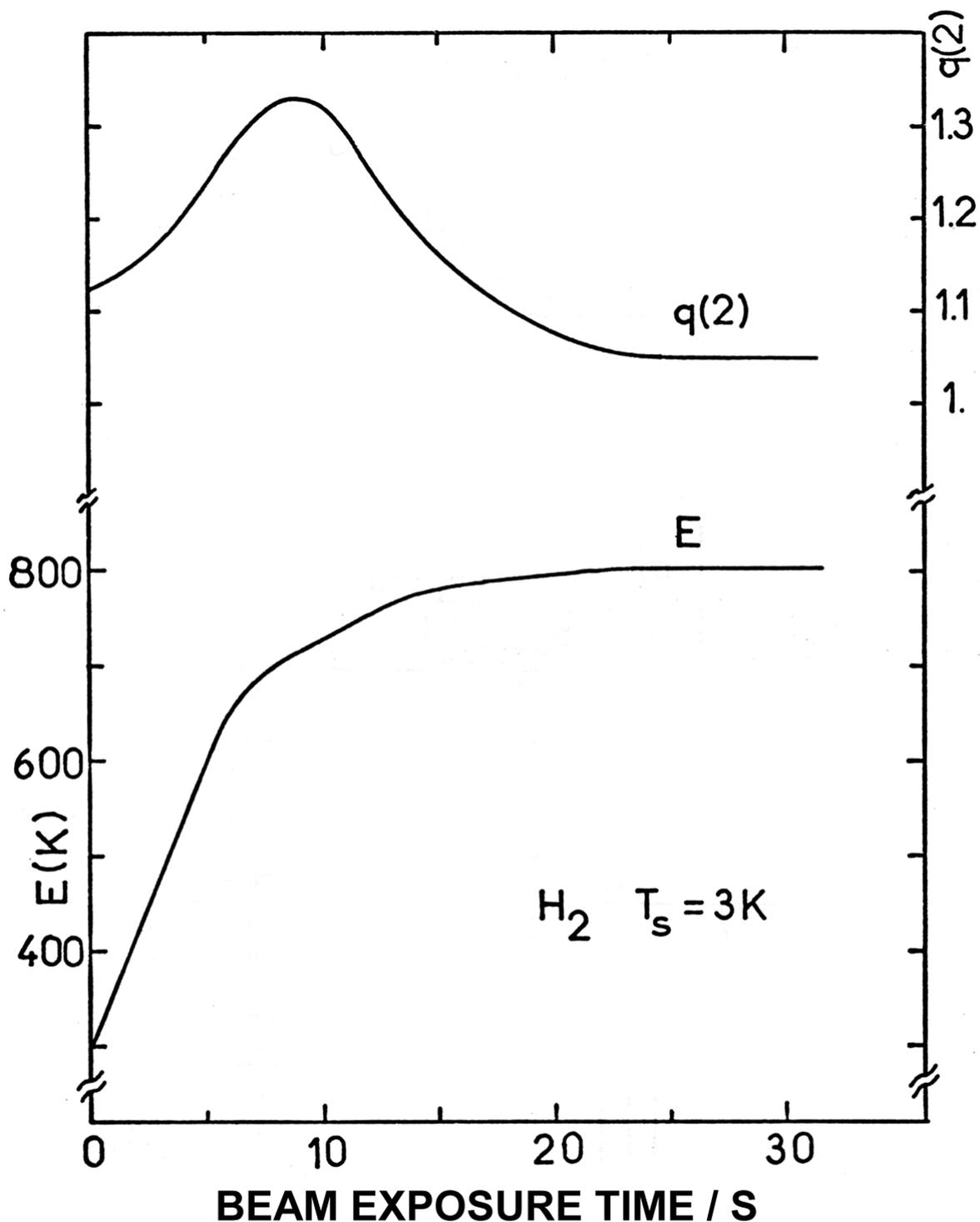

Figure 4:

Simultaneous bolometer- and mass spectrometer signals as a function of time of exposure to a $H_2$ beam with a flux (averaged over the chopping period) on the bolometer surface of $2.0 \times 10^{14}$ $cm^{-2}s^{-1}$. At t=0 the bolometer is free of adsorbed hydrogen. The bolometer signal E is expressed as the energy transferred per incident molecule. The mass spectrometer signal q(2) is normalised to the signal one would obtain upon elastic reflection of the beam. Th incidence and QMS detection angles are 45°. The bolometer temperature is 3 K. (from ref. [7])

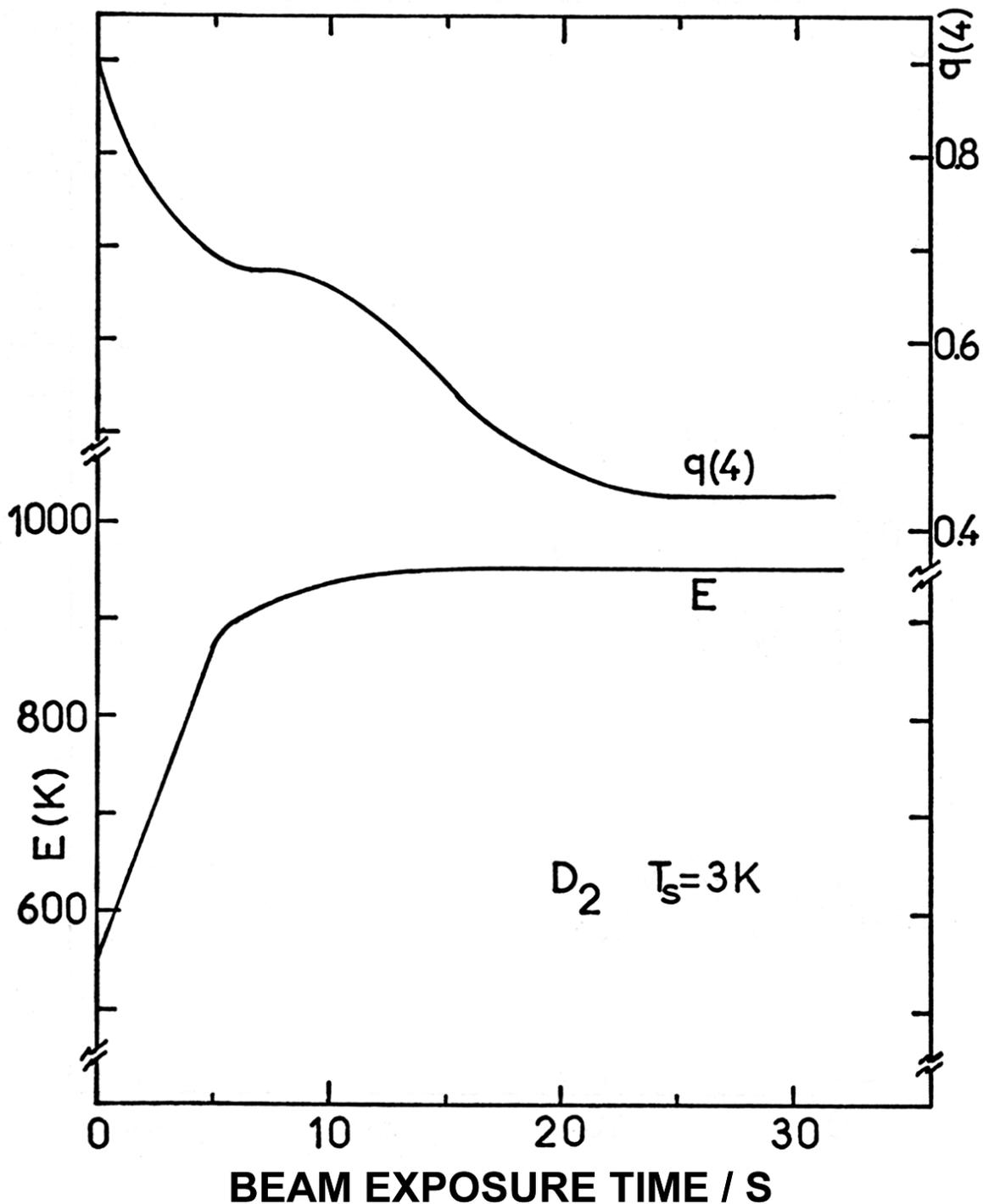

Figure 5:

Simultaneous bolometer- and mass spectrometer signals as a function of time of exposure to a D$_2$ beam with a flux (averaged over the chopping period) on the bolometer surface of $1.2 \times 10^{14}$ cm$^{-2}$s$^{-1}$. At t=0 the bolometer is free of adsorbed hydrogen. The bolometer signal E is expressed as the energy transferred per incident molecule. The mass spectrometer signal q(4) is normalised to the signal one would obtain upon elastic reflection of the beam. The incidence and QMS detection angles are 45°. The bolometer temperature is 3 K. (from ref. [7])

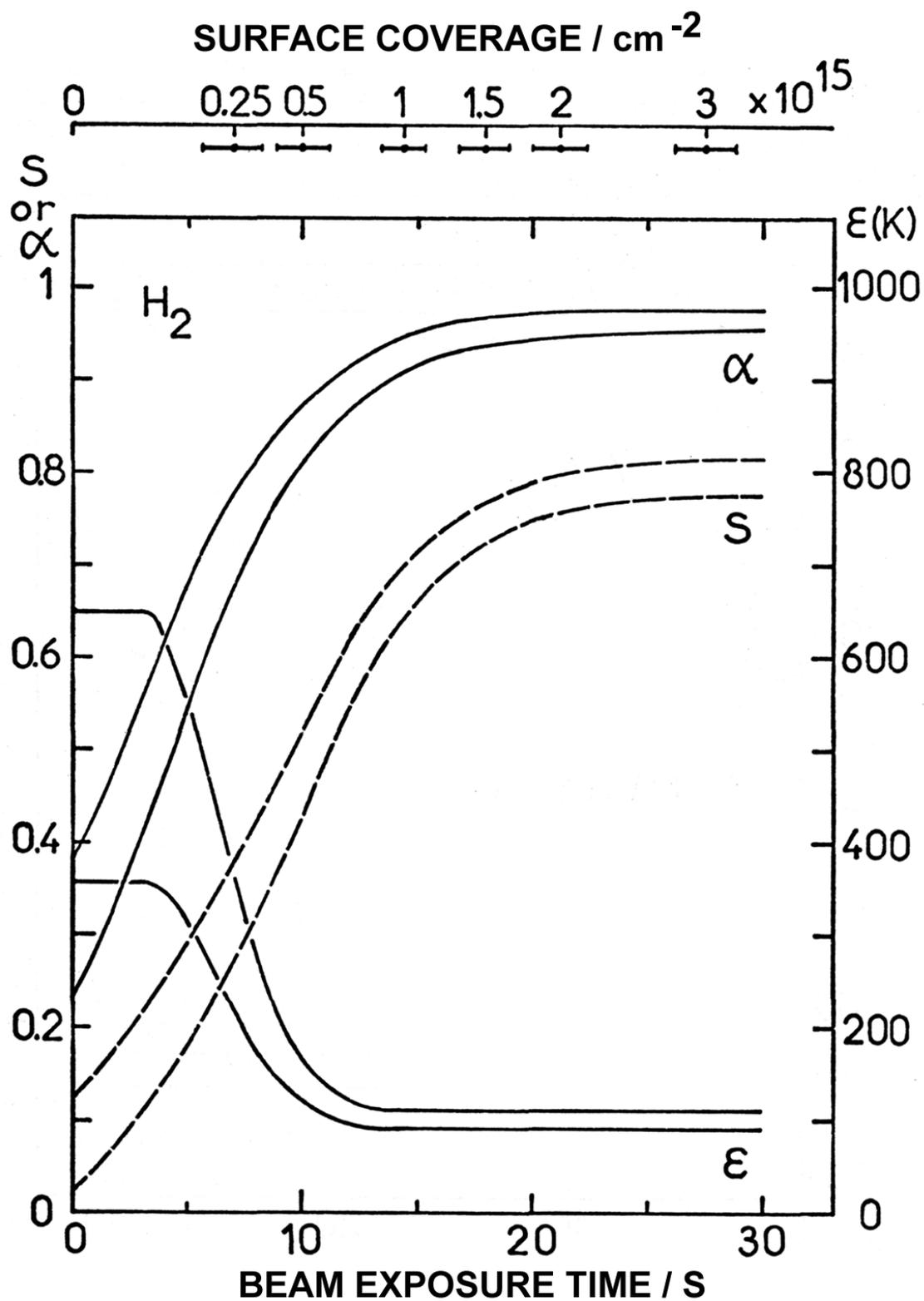

Figure 6:

Upper and lower bounds to the sticking coefficient S and the accommodation coefficient $\alpha$ obtained from the $H_2$ data of fig. 4 when the binding energy $\varepsilon$ is allowed to vary within the limits shown. The upper coverage scale was obtained from the incident beam flux I by integrating the product [I.S(t)]. (from ref. [7])

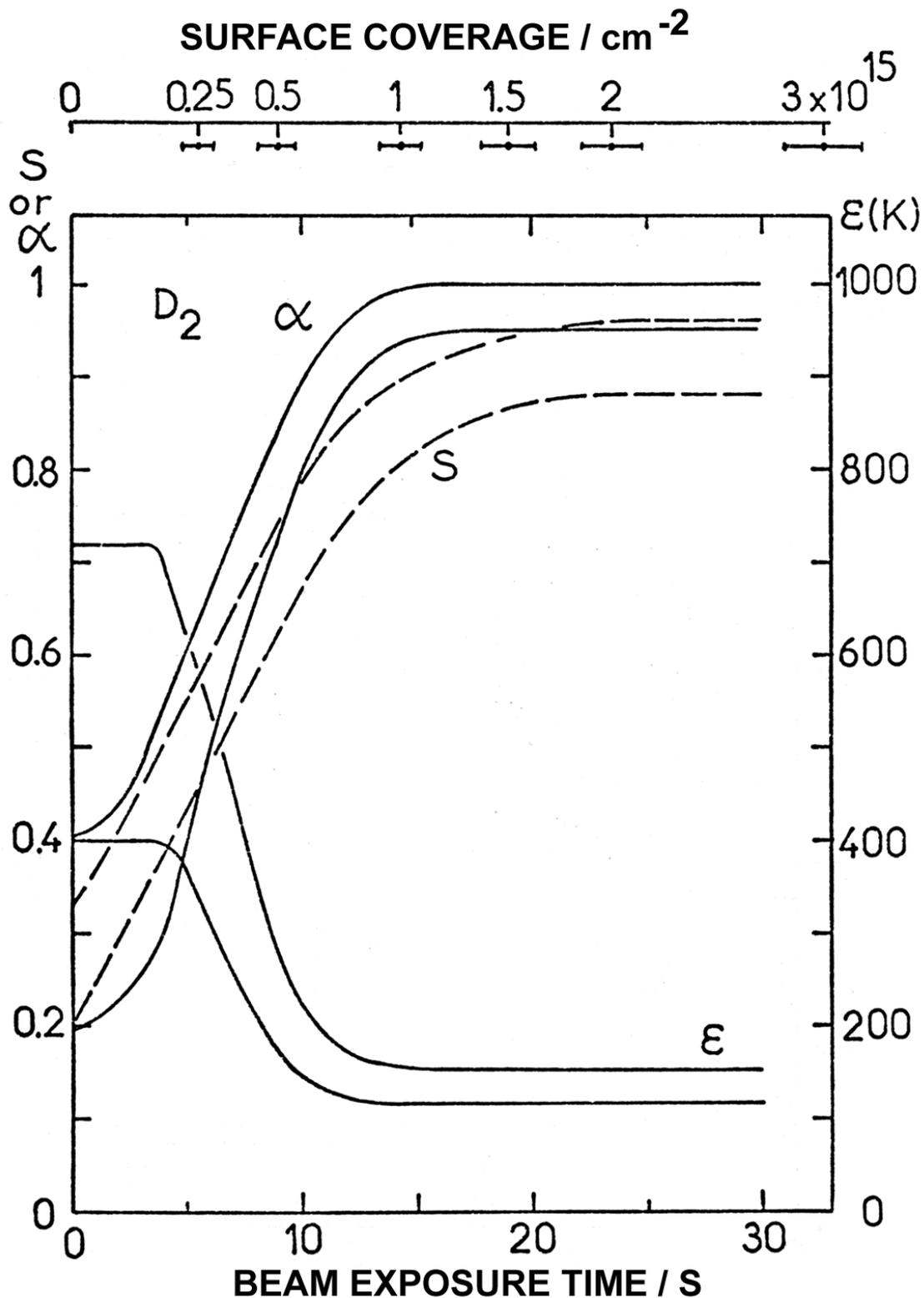

Figure 7:

Upper and lower bounds to the sticking coefficient S and the accommodation coefficient α obtained from the $D_2$ data of fig. 5 when the binding energy ε is allowed to vary within the limits shown. The upper coverage scale was obtained from the incident beam flux I by integrating the product [I.S(t)]. (from ref. [7])

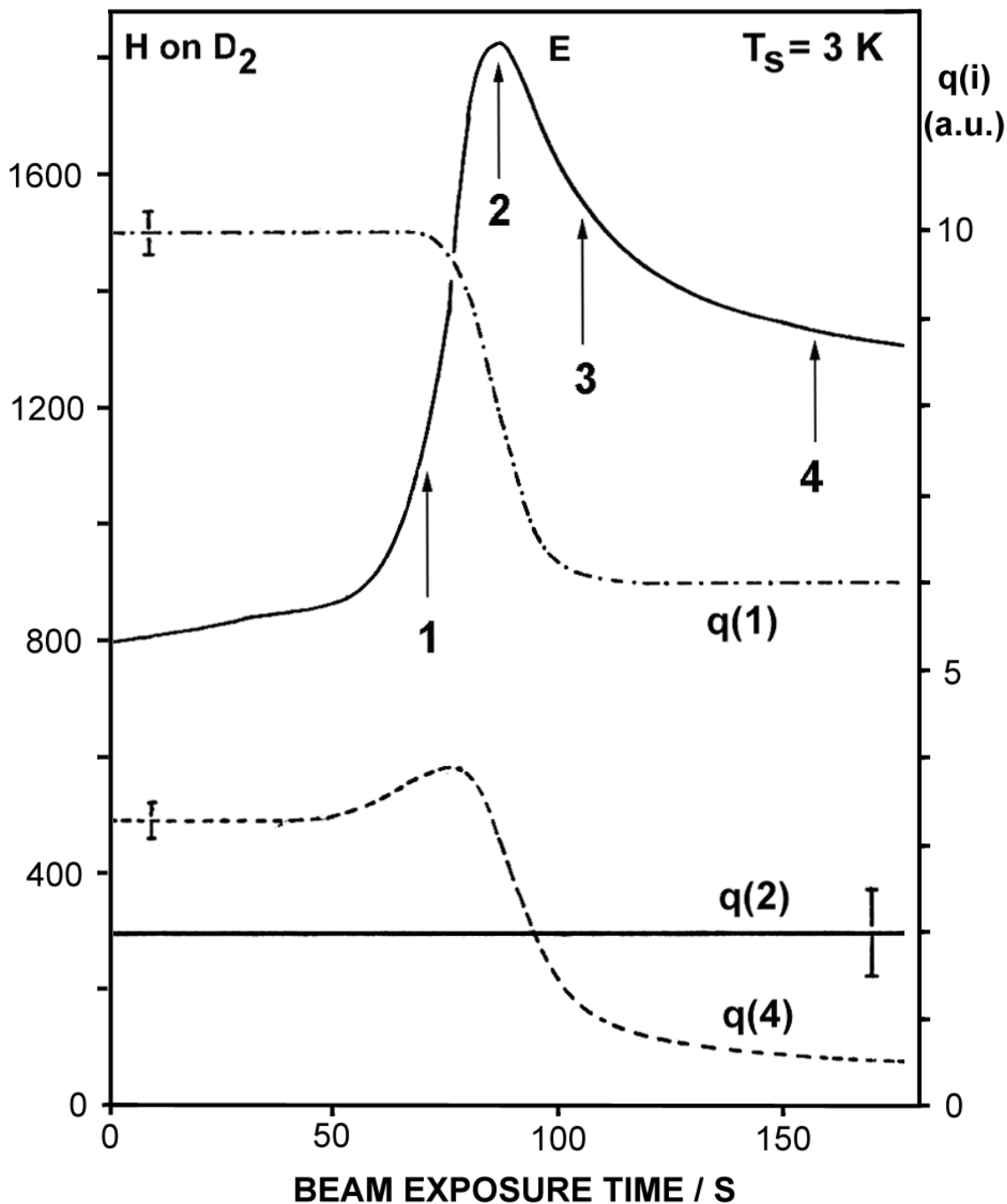

Figure 8:

Simultaneous bolometer- and mass spectrometer signals upon impact of a hydrogen atom beam with a flux (averaged over the beam chopping period) of $3.8 \times 10^{14}$ cm$^{-2}$s$^{-1}$ on a surface initially saturated with $D_2$, corresponding to a coverage of about $3 \times 10^{15}$cm$^{-2}$. The bolometer signal E is expressed as energy transferred per incident atom. The QMS signals q(i) are in arbitrary units. When the bolometer signal starts rising sharply (arrow 1), the H beam has reduced the $D_2$ coverage to about $9 \times 10^{14}$cm$^{-2}$. At the maximum (arrow 2) it is about $2 \times 10^{14}$cm$^{-2}$, decreasing further to some 1.5 and $0.4 \times 10^{14}$cm$^{-2}$ at the times indicating by arrows 3 and 4. (data from ref. [6])

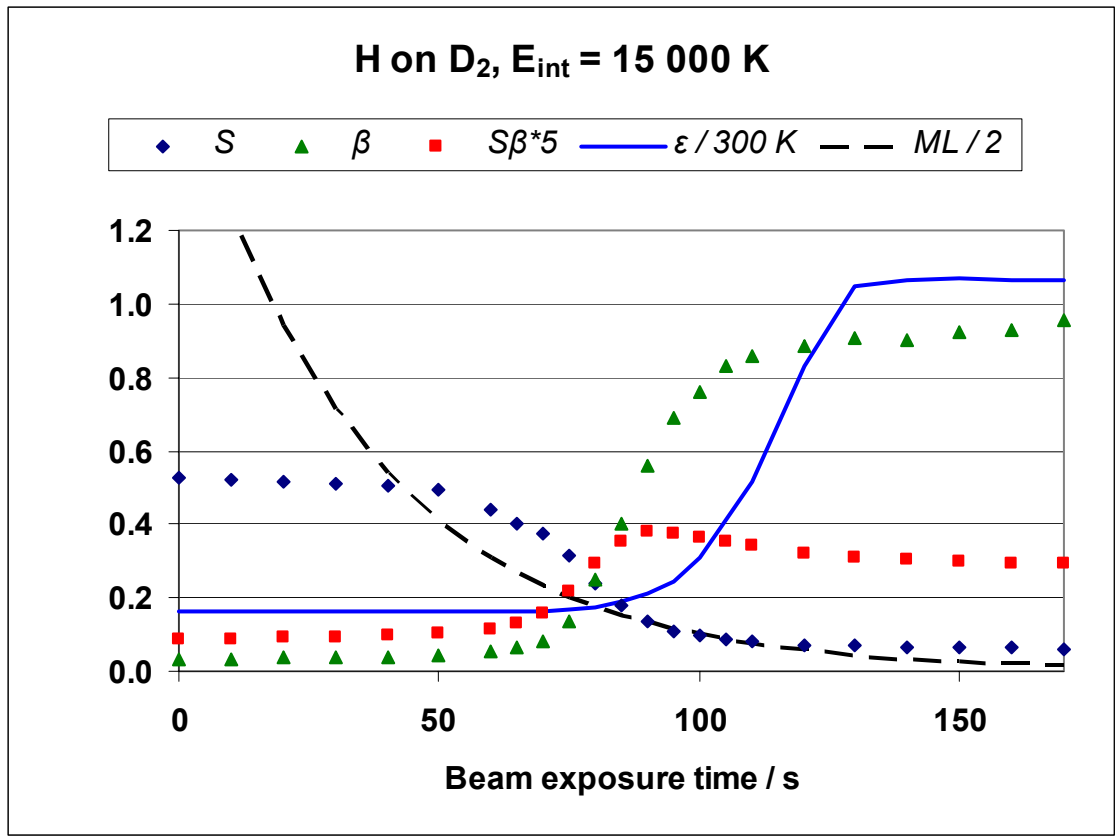

Fig. 9:

Time evolution of $S$, $\beta$ and the recombination probability ($S.\beta$) deduced from the data of fig. 8 under the assumption that an energy of 15000 K is stored as internal energy of the $H_2$ formed by recombination. Comparison with fig. 8 shows that the maximum in the bolometer signal corresponds to the maximum in the product ($S.\beta$). It results with opposing variations in $S$ and $\beta$, as the H beam reduces $D_2$ coverage from its initial value of about $3 \times 10^{15} cm^{-2}$ to $2 \times 10^{14} cm^{-2}$ at the peak after 90 s exposure, and to about $0.4 \times 10^{14} cm^{-2}$ after 160 s exposure. Note that ($S.\beta$) has been multiplied by 5 in the figure. The surface coverage by $D_2$, devided by $2 \times 10^{15} cm^{-2}$ (2 ML), is indicated by the dashed line, and the H binding energy (devided by 300 K) is indicated by the solid line (data from ref. [6]).

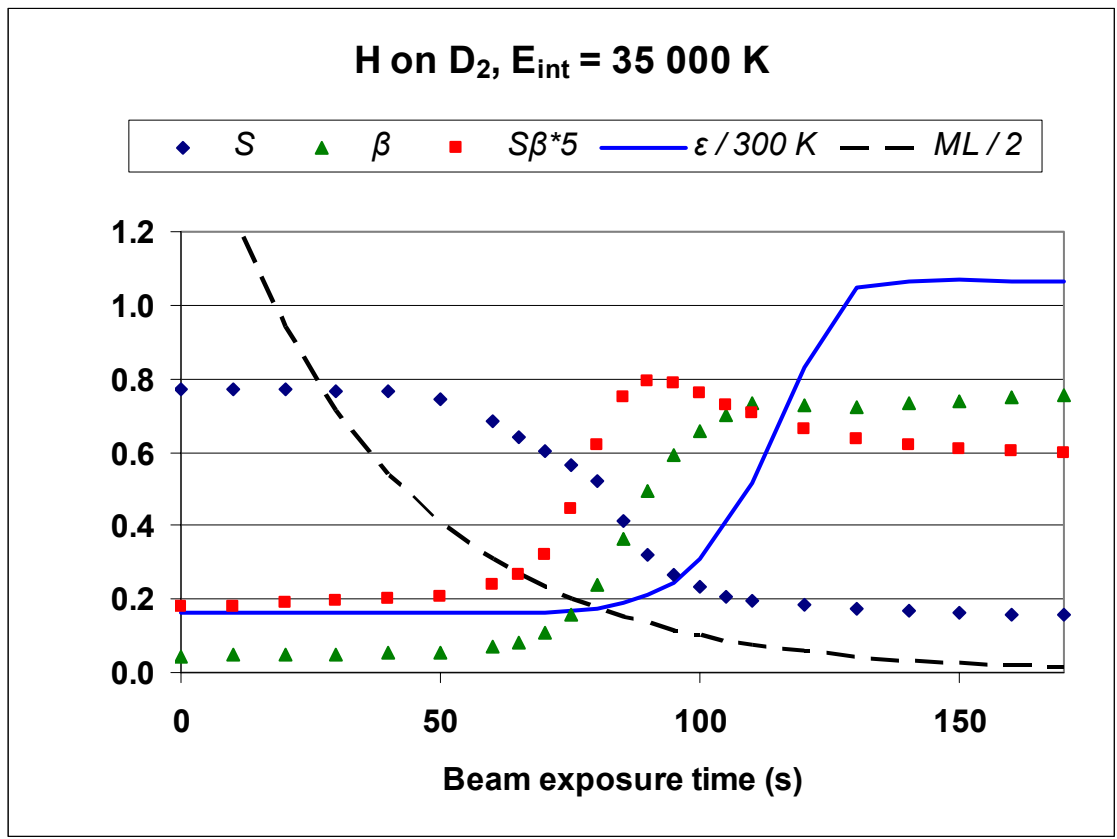

Fig.10:

Time evolution of $S$, $\beta$ and the recombination probability $(S.\beta)$ deduced from the data of fig. 8 under the assumption that an energy of 35000 K is stored as internal energy of the $H_2$ formed by recombination. As can be seen from a comparison with fig. 9, adopting a higher value for the internal energy implies higher values for $S$ and $(S.\beta)$. The $D_2$ coverage diminishes from its initial value of about $3 \times 10^{15} cm^{-2}$ to $2 \times 10^{14} cm^{-2}$ at the peak after 90 s exposure, and to about $0.4 \times 10^{14} cm^{-2}$ after 160 s exposure. Note that $(S.\beta)$ has been multiplied by 5 in the figure. The surface coverage by $D_2$, devided by $2 \times 10^{15} cm^{-2}$ (2 ML), is indicated by the dashed line, and the H binding energy (devided by 300 K) is indicated by the solid line (data from ref. [6]).

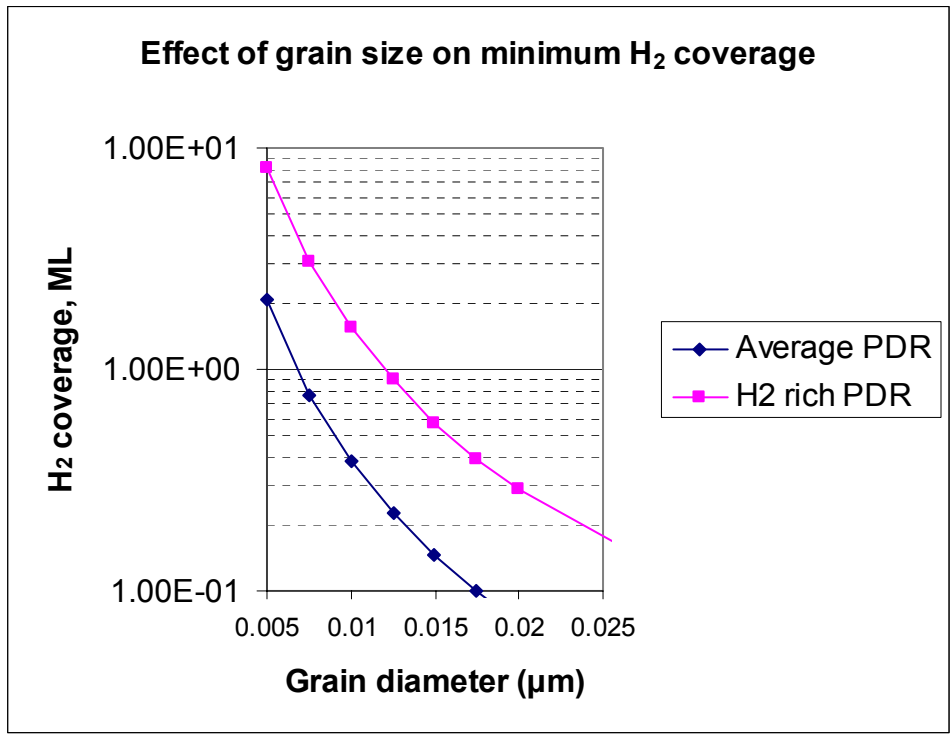

Figure 11:

Lower limits to molecular hydrogen coverage of grains in a PDR cloud as a function of grain diameter. These values were calculated by assuming a binding energy 1.2 times larger for $H_2$ than the lowest value for H compatible with the minimum residence times evaluated as discussed in sect. 1, but with $S(H) = 0.6$ rather than $S(H) = 1$. Molecular density for the "average" PDR was assumed to be $1 \times 10^4$ cm$^{-3}$, and that for the $H_2$-rich cloud was $4 \times 10^4$ cm$^{-3}$, at a gas temperature of 300 K, a surface temperature of 10 K and a H-atom density of $5 \times 10^4$ cm$^{-3}$. The molecular sticking coefficient was taken to be $S(H_2) = 0.7$.